\documentclass[12pt,epsfig]{article}
\usepackage[dvips]{graphicx}
\usepackage{amsfonts}
\usepackage{amsmath}
\usepackage{amssymb}
\usepackage{color}

\def\*{{\phantom *}}

\newtheorem{postulate}{Hypothesis}[section]  
\newcommand{\bhypo}{\begin{postulate} \it}
\newcommand{\ehypo}{\end{postulate}}

\newtheorem{theorem}{Theorem}[section]  
\newcommand{\btheo}{\begin{theorem} \it}
\newcommand{\etheo}{\end{theorem}}

\newtheorem{proposition}{Proposition}[section]  
\newcommand{\bprop}{\begin{proposition}\it}
\newcommand{\eprop}{\end{proposition}}

\newtheorem{corollary}{Corollary}[section]  
\newcommand{\bcorol}{\begin{corollary} \it}
\newcommand{\ecorol}{\end{corollary}}

\newtheorem{lemma}{Lemma}[section]  
\newcommand{\blem}{\begin{lemma}\it}
\newcommand{\elem}{\end{lemma}}

\newtheorem{remark}{Remark}[section]   
\def\brem{\begin{remark} \rm }
\def\erem{\end{remark}}

\newtheorem{naming}{Definition}[section]   
\newcommand{\bdefi}{\begin{naming} \rm }
\newcommand{\edefi}{\end{naming}}

\begin{document}

\phantom{.} {\qquad \hfill \textit{\textbf{Dedicated to John T.Lewis}}}
\vskip 1.5cm
\begin{center}
{\Large{\bf {Bose-Einstein Condensation in the Luttinger-Sy Model}}}
\linebreak
\vskip 1cm

\setcounter{footnote}{0}
\renewcommand{\thefootnote}{\arabic{footnote}}

\vspace{1cm} {\bf Olivier LENOBLE\footnote{University of California,
Irvine - {{email: lenoble@math.uci.edu}}} and Valentin A.
ZAGREBNOV\footnote{Universit\'e de la M\'editerran\'ee
(Aix-Marseille II) - email : zagrebnov@cpt.univ-mrs.fr}}

\vspace{0.7 cm}

\textbf{University of California at Irvine, Department of Mathematics, \\
92697-38 Irvine, California, USA$^{1}$}

\vspace{0.2 cm}

\textbf{Universit\'e de la M\'editerran\'ee and Centre de Physique
Th\'eorique, \\ Luminy-Case 907, \\
13288 Marseille, Cedex 09, France$^{2}$}

\vspace{1.5cm}
\end{center}

\begin{abstract}
\noindent We present a rigorous study of the Bose-Einstein
condensation in the \textit{Luttinger-Sy} model. We prove the
existence of the condensation in this one-dimensional model of the
perfect boson gas placed in the Poisson random potential of singular
point impurities. To tackle the off-diagonal long-range order we
calculate explicitly the corresponding space-averaged one-body
reduced density matrix. We show that mathematical mechanism of the
Bose-Einstein condensation in this random model is similar to
condensation in a one-dimensional \textit{nonrandom hierarchical}
model of scaled intervals. For the Luttinger-Sy model we prove the
\textit{Kac-Luttinger conjecture}, i.e., that this model manifests a
\textit{type} I BEC localized in a single "largest" interval of
\textit{logarithmic} size.

\end{abstract}
\hspace*{2.2cm} Lifshitz tail, Kac-Luttinger conjecture
\textit{and}\\
\textbf{AMS:} 82B10, 82B23, 81V70

\newpage

\section{Introduction}
\setcounter{equation}{0}
\renewcommand{\theequation}{\arabic{section}.\arabic{equation}} 

\textbf{1.1} In our recent paper \cite{LPZ}, we presented some
general mathematical results concerning the existence of the
\textit{Bose-Einstein condensation} (BEC) of the \textit{Perfect
Bose-Gas} (PBG) placed in a semi-bounded from below homogeneous
ergodic random external potential (random impurities). There we show
that for the infinite-volume one-particle Schr\"{o}dinger
operator{,} a generic \textit{Lifshitz tail} behaviour of density of
states near the lower edge of the spectrum reduces the
\textit{critical dimensionality} of the BEC for PBG from
dimensionality $d= 2+\varepsilon$ to $d=1$. Therefore, the
randomness \textit{enhances} the BEC, and moreover, it is shown to
be stable with respect to the \textit{mean-field} particle
interaction \cite{Le}.

To tackle the corresponding \textit{Off-Diagonal Long-Range Order}
(ODLRO){,} we introduced in \cite{LPZ} a concept of the
\textit{space-averaged} one-body reduced density matrix. In spite of
a rather accurate estimate of this matrix \textit{out} of the BEC
domain, that shows an enhancement of the ODLRO exponential
\textit{decay} due to impurities, we did not obtain in \cite{LPZ}
any sound estimate for this order \textit{in} the BEC domain.

\noindent \textbf{1.2} The aim of this paper is to present a
rigorous study of a particular case of {a} one-dimensional PBG model
in homogeneous ergodic non-negative random potential induced by the
Poisson distributed \textit{singular point} impurities (the
\textit{Luttinger-Sy} model \cite{Lutt-Sy-1,Lutt-Sy-2}). We
show that this model allows a rigorous mathematical approach to
condensation and that one can compute explicitly some of
thermodynamical quantities even in the BEC domain. This concerns in
particular the ODLRO behaviour of the two-point correlation function
(space-averaged one-body reduced density matrix) in the condensation
regime.

Notice that the first study of possible modifications of
$d=3$ dimensional BEC in the PBG caused by repulsive finite-range
impurities goes back to Kac and Luttinger \cite{KL1,KL2}.
They predicted an enhancement of $d=3$ BEC by indication that due to impurities there
is decreasing of the critical density , but did not discuss a modification
of the critical dimensionality. {They also mentioned  a puzzling
question} about the nature of the established BEC. For example, they
\textit{conjectured} that this condensate occurs as a
macroscopic occupation of only the ground-state: \textit{type I}
BEC. We prove this conjecture in the case of the Luttinger-Sy model,
see discussion in Section 6. We show that the nature of BEC in this model
is close to what is known as the "\textit{Bose-glass}", since it may be
localized by the random potential. This is of interest for example in
experiments with liquid $^4He$ in random environments like Aerogel and Vycor
glasses, \cite{FWGF,KT}.

On the other hand, the nature and behaviour of the \textit{lattice}
BEC may be quite different. First of all, the lattice Laplacian and
the on-site \textit{Bose-Hubbard} particle {repulsion} produces a
coexistence of the BEC (\textit{superfluidity}) and the \textit{Mott
insulating phase} as well as domains of \textit{incompressibility},
see e.g. a very complete review \cite{U}. Adding disorder makes the
corresponding models much more complicated. The physical arguments
show that the randomness may  \textit{suppress} the BEC
(superfluidity) as well as the Mott phase in favour of the localized
\textit{Bose-glass} phase, but this is very sensitive to the choice
of the random distribution, for some recent rigorous results see
\cite{DPZ}.

\noindent \textbf{1.3} The paper is organized as follows. In Section
2 we recall definition of the Luttinger-Sy model and some of its
properties. We prove the \textit{self-averaging} of  corresponding
\textit{integrated density of states} in Section 3 and we calculate
it explicitly. In Section 4 we prove that the established integrated
density of states implies the existence of \textit{generalized} BEC
in the case of PBG.

Our main results are collected in Sections 5 and 6. There we recall the
notion of the \textit{space-averaged} one-body reduced density
matrix and we prove that {it has an \textit{almost sure} nonrandom
thermodynamic} limit (self-averaging), which can be calculated
explicitly for all values of particle density. We also prove that
randomness \textit{enhances} decay of the two-point correlation
function. In particular we show that it keeps this decay always
\textit{exponential}, even in the presence of BEC. We found that in
the latter case the ODLRO is non-zero and that it coincides with the
condensate density.

The properties of the BEC are discussed in concluding Section 6.
First, we analyze the critical density dependence on the amplitude
of the repulsive Poisson point impurities. Notice that for the
Luttinger-Sy model the singular point impurities mean that this
amplitude is infinite. Next, we study the problem of the condensate
nature and its \textit{localization}. To elucidate this point we
invented a \textit{hierarchical} one-dimensional \textit{nonrandom}
model, which mimics in {a} certain sense the (random) Luttinger-Sy
model. We show that this hierarchical model can manifest different
\textit{types }(I,II and III) of \textit{generalized} van den
Berg-Lewis-Pul\'{e} condensations \cite{VDB-Lew-Pu} localized in one, several
or infinite number of (\textit{infinite}) intervals of \textit{logarithmic} sizes. To
discriminate between these options, i.e. to prove or disprove the
Kac-Luttinger \textit{conjecture},
one has to have a quite detailed information about the energy level spacing in random
intervals generated by the Poisson impurity positions. We prove this conjecture,
i.e., that \textit{type} I BEC in the Luttinger-Sy model is localized in a single
"largest" (i.e. infinite) interval of the \textit{logarithmic} size.

\section{The Luttinger-Sy Model}
\setcounter{equation}{0}
\renewcommand{\theequation}{\arabic{section}.\arabic{equation}} 

In the framework of general setting this model corresponds to the following
one-dimensional ($d=1$) single-particle \textit{random}
Schr\"{o}dinger operator in the Hilbert space $\mathcal{H}=L^2(\mathbb{R})$:\\
\textbf{2.1} Consider a random (measurable) potential $v^{(\cdot
)}(\cdot ):\Omega \times \mathbb{R }\rightarrow
\mathbb{R},(\omega,x)\mapsto v^{\omega }(x)$, which is a random
field on a probability space $(\Omega ,\mathcal{F},\mathbb{P})$,
with the properties:

(a) $v^{\omega }$ is homogeneous and ergodic with respect to the
group $\left\{ \tau _{x}\right\} _{x\in \mathbb{R}}$ of
probability-preserving translations on $(\Omega ,\mathcal{F},
\mathbb{P})$;

(b) $v^{\omega }$ is non-negative and $\inf_{x\in
\mathbb{R}^{d}}\left\{ v^{\omega }(x)\right\} = 0$.

\noindent By $\mathbb{E}\left\{ \cdot \right\} :=\int_{\Omega
}\mathbb{P}(d\omega )\left\{ \cdot \right\} $ we denote the
expectation with respect to the probability measure in $(\Omega
,\mathcal{F},\mathbb{P})$. Then the \textit{random}
Schr\"{o}dinger operator corresponding to the potential $v^{\omega}$ is
a family of random operators $\left\{ h^{\omega }\right\}_{\omega \in \Omega}$ :
\begin{equation}\label{Schr1}
h^{\omega }:= t+v^{\omega },
\end{equation}
where $t := (-\Delta /2)$ is the \textit{free} one-particle
Hamiltonian, i.e., a unique self-adjoint extension of the operator:
$-\Delta /2$, with domain in $L^{2}(\mathbb{R})$.

Notice that assumptions (a) and (b) guarantee
that there exists a subset $\Omega _{0}\subset \mathcal{F}$ with $\mathbb{P}%
(\Omega _{0})=1$ such that operator (\ref{Schr1}) is
\textit{essentially} self-adjoint on domain $\mathcal{C}_{0}^{\infty
}(\mathbb{R})$ for every $\omega \in
\Omega _{0}$ (see e.g. \cite{PastFig} Ch.I.2).\\
\textbf{2.2} Let $u(x)\geq 0,\,x\in \mathbb{R}$, be continuous
function with a \textit{compact} support. We call it a
(\textit{repulsive}) single-impurity potential. Let $\left\{ \nu
_{\lambda }^{\omega }(dx)\right\} _{\omega \in \Omega }$ be
\textit{random} Poisson measure on $\mathbb{R}$ with intensity
$\lambda >0$ :
\begin{equation}
\mathbb{P}\left( \left\{ \omega \in \Omega :\nu _{\lambda
}^{\omega }(\Lambda )=n\right\} \right) =\frac{\left( \lambda
\left\vert \Lambda \right\vert
\right) ^{n}}{n!}e^{-\lambda \left\vert \Lambda \right\vert }\, \ , \ \ n\in \mathbb{N%
}_{0}=\mathbb{N}\cup \left\{ 0\right\} ,  \label{Ran-Poiss-mes}
\end{equation}%
for any bounded Borel set $\Lambda \subset \mathbb{R}$. Then the
non-negative random potential $v^{\omega}$ generated by the Poisson distributed
local impurities has realizations
\begin{equation}
v^{\omega }(x):=\int_{\mathbb{R}}\nu _{\lambda }^{\omega
}(dy)u(x-y)=\sum_{x_{j}^\omega\in X^\omega}u(x-x_{j}^{\omega }).
\label{Ran-Poiss-pot}
\end{equation}%
Here the random set $X^\omega$ corresponds to impurity positions
$X^\omega = \left\{ x_{j}^{\omega }\right\}_j \subset \mathbb{R}$,
which are the atoms of the random Poisson measure, i.e.,  $card \
\{X^{\omega}\upharpoonright \Lambda\}=
\nu_{\lambda}^{\omega}(\Lambda)$ equals to the number of impurities
in the set $\Lambda$. Since the expectation
$\mathbb{E}\left(\nu_{\lambda}^{\omega}(\Lambda) \right)=
\lambda\left|\Lambda\right|$, the parameter $\lambda$ coincides with
impurities \textit{concentration} on the axe $\mathbb{R} $.
\begin{remark}\label{propert-Poisson-example-1} {The random potential}
(\ref{Ran-Poiss-pot}) is obviously homogeneous and ergodic (even
strongly mixing), i.e. it verifies the conditions (a) and
(b). Moreover, see \cite{PastFig} Ch.II.5, we have that:\\
- There exists a nonrandom measure $d\mathcal{N}(E)$ on $\mathbb{R}
$ such that
\begin{equation}
d\mathcal{N}(E):=\mathbb{E}\left\{ \mathcal{E}_{h^{\omega
}}(dE;0,0)\right\} . \label{lim-meas2}
\end{equation}%
Here $\mathcal{E}_{h^{\omega }}(dE;x,y)$ is the kernel of the
spectral decomposition measure corresponding to random
Schr\"{o}dinger operator $h^{\omega }$. The spectrum
$\sigma(h^{\omega})$ of $h^{\omega}$ is {almost-surely}
{\rm{(}a.s.\rm{)}} nonrandom and it coincides with the support of
$\mathcal{N}$: $
\;\sigma(h^{\omega}) = \mathrm{supp}\, \mathcal{N}.$\\
- For repulsive impurities with compact support and for Poisson
distribution{,} the a.s nonrandom spectrum $\sigma (h^{\omega }) =
\mathbb{R}_{+}$. Thus the lower edge of the spectrum
$\inf\left\{\sigma (h^{\omega})\right\}=0$, i.e. it
coincides with the lower edge of the spectrum of {the} nonrandom operator $t$, see (\ref{Schr1}).\\
- In one-dimensional case the asymptotic behaviour of the
integrated density of states $\mathcal{N}(E):= \mathcal{N}((-\infty, E])$ as $E\downarrow 0$
has the form (the \textit{Lifshitz tail}):
\begin{equation}
\ln \mathcal{N}(E)\sim -\lambda \left( \frac{c_d}{E}\right)
^{d/2}\,,  \ E\downarrow 0 \ ,   \label{Lifsh-tail}
\end{equation}%
for $c_d >0$. Recall that in the nonrandom case $v^\omega =0$ one
obtains: $\mathcal{N}(E)\sim E^{d/2},\,\,E\downarrow 0$.
\end{remark}
\textbf{2.3} Luttinger and Sy defined their $d=1$ model
\cite{Lutt-Sy-1} restricting the  single-impurity potential to the
\textit{point} $\delta$-potential with amplitude $a>0$. In fact this choice
(even for more general case of random $\left\{a_j\right\}_j$) goes
back to Frish and Lloyd \cite{FrLl}. Then the corresponding random
potential (\ref{Ran-Poiss-pot}) takes the form:
\begin{equation}
v_{a}^{\omega }(x):=\int_{\mathbb{R}}\nu _{\lambda}^{\omega
}(dy)a\delta (x-y)=a\sum_{x_{j}^\omega\in X^\omega}\delta
(x-x_{j}^{\omega }). \label{ran-delta1}
\end{equation}
Now the self-adjoint one-particle random Schr\"{o}dinger operator
\begin{equation}\label{Schr2}
h_{a}^{\omega}:=t \dotplus v_{a}^{\omega},
\end{equation}
can be defined in the sense of the sum of quadratic forms. In
spite of a singular nature of this random potential, by standard limiting
arguments \cite{PastFig} it inherits the properties quoted in
Remark \ref{propert-Poisson-example-1}.\\
\textbf{2.4} Moreover, the same arguments \cite{PastFig} are applied
to define a \textit{strong resolvent} (s.r.) limit of Hamiltonians
(\ref{Schr2}), when $a \rightarrow +\infty$, which is the
\textit{last step} in definition of the Luttinger-Sy model,
\cite{Lutt-Sy-1}. This limit gives the self-adjoint
(\textit{Friedrichs}) extension of symmetric operator $t_0
=-\Delta/2$ with domain $dom(t_0)= \left\{f\in \mathcal{H}: f\in
\mathcal{C}_{0}^{\infty}(\mathbb{R}\setminus X^{\omega})\right\}$.
For any $\omega\in \Omega$ we denote this extension by
\begin{equation}\label{s.r.}
h_{D}^{\omega}:= s.r. \lim_{a \rightarrow +\infty} h_{a}^{\omega} \ .
\end{equation}
Since for any $\omega\in\Omega$ the set $X^{\omega}$ can be
\textit{ordered}: $X^\omega = \left\{ x_{j}^{\omega }\right\}_j$ ,
it generates a set of intervals $\left\{I_{j}^{\omega}:=
(x_{j-1}^{\omega},x_{j}^{\omega})\right\}_j$ of lengths
$\left\{L_{j}^{\omega}:= x_{j}^{\omega} -
x_{j-1}^{\omega}\right\}_j$ . Then one can decompose {the} Hilbert
space $\mathcal{H}=L^2(\mathbb{R})$ into (random) direct orthogonal
sum:
\begin{equation}\label{space-orth-sum}
\mathcal{H} = \bigoplus_{j}\mathcal{H}_{j} \,,  \,\,\,\,
\mathcal{H}_{j}:= L^2(I_{j}^{\omega}).
\end{equation}
Correspondingly, let $h_D(I_{j}^{\omega})$ denote the \textit{Friedrichs} extension of operator
$t_0 = -\Delta/2$ with domain ${\rm{dom}}(t_0) = \left\{f\in L^{2}(I_{j}^{\omega}):
f\in \mathcal{C}_{0}^{\infty}(I_{j}^{\omega})\right\}$:
\begin{eqnarray}\label{ham-Dirich}
&&(h_D(I_{j}^{\omega})f)(x):= -\frac{1}{2}(\Delta f)(x)\,, \\
&&f\in {\rm{dom}}(h_D(I_{j}^{\omega}))= \left\{f \in W^{2}_2(I_{j}^{\omega}):
f(x_{j-1}^{\omega}) = f(x_{j}^{\omega})= 0\right\} , \nonumber
\end{eqnarray}
where $W^{2}_2$ denotes the corresponding Sobolev space. Then we get decompositions of the one-particle
Luttinger-Sy Hamiltonian:
\begin{equation}\label{ham-orthg-sum}
h_{D}^{\omega} =
\bigoplus_{j}h_D(I_{j}^{\omega}) \ , \ \ \omega\in\Omega  \ ,
\end{equation}
with domain
\begin{equation}\label{dom-orth-sum}
{\rm{dom}}(h_{D}^{\omega})= \bigoplus_{j} {\rm{dom}}(h_D(I_{j}^{\omega}))\subset \mathcal{H} \ ,
\end{equation}
into random disjoint free Schr\"{o}dinger operators
$\left\{h_D(I_{j}^{\omega})\right\}_{j , \omega}$ with
\textit{Dirichlet} boundary conditions at the end-points of {the}
intervals $\left\{I_{j}^{\omega}\right\}_j$. The corresponding
eigenfunctions have the form:
\begin{equation}\label{eigenfunctions-orth-space}
\Psi_{s_j,D}^{\omega}(x) = (0,0,\ldots,\psi_{j,s_j}^{\omega}(x),0,\ldots) \ ,
\end{equation}
with eigenvalues $\left\{E_{s_j}(L_j^\omega)\right\}_{s_j}$:
\begin{equation}\label{eigenvalues-orth-space}
h_{D}^{\omega} \ \Psi_{s_j,D}^{\omega}= E_{s_j}(L_j^\omega) \ \Psi_{s_j,D}^{\omega} \ .
\end{equation}
\begin{remark}\label{spectral-L-S-1}
For a given realization $\omega \in \Omega$ the spectrum of {the}
random operator (\ref{ham-Dirich}) is explicitly defined by
non-degenerate eigenvalues
\begin{equation}\label{eigenvalues}
\sigma(h_{D}(I_{j}^{\omega}))=
\left\{E_{s_j}(L_j^\omega)=\frac{1}{2}\frac{\pi^2s_j^2}{(L_j^\omega)^2}\right\}_{s_j=1}^{\infty}\,\,,
\end{equation}
with {the} corresponding eigenfunctions
\begin{equation}\label{eigenfunctions}
\psi_{j,s_j}^{\omega}(x)=\mathbb{I}_{I_{j}^{\omega}}(x)\ \sqrt{\frac{2}{L_j^\omega}}\sin(\frac{\pi
s_j} {L_j^\omega}(x-x_{j-1}^\omega)) \ .
\end{equation}
Here $\mathbb{I}_{I_{j}^{\omega}}(x)$ is the characteristic
function of the interval ${I_{j}^{\omega}}$. By consequence, the
spectrum of the Luttinger-Sy Hamiltonian (\ref{ham-orthg-sum}) is
the union of (\ref{eigenvalues})
\begin{equation}\label{spect-L-S}
\sigma(h_{D}^{\omega})= \bigcup_j \sigma(h_{D}(I_{j}^{\omega})).
\end{equation}
\end{remark}

By virtue of Remark \ref{propert-Poisson-example-1} this spectrum is
a.s. \textit{nonrandom}, and it coincides with support of the integrated
density of states $\mathcal{N}$. Moreover, in the case of the
Luttinger-Sy Hamiltonian (\ref{ham-orthg-sum}) it is known
explicitly since \cite{Lutt-Sy-1}. But the rigorous study and in
particular  the concept of "self-averaging", which ensures this
nonrandom property, are due to \cite{LGP}.

\section{Self-averaging of the integrated density of states}
\setcounter{equation}{0}
\renewcommand{\theequation}{\arabic{section}.\arabic{equation}} 

For the reader {convenience} we recall in this section some arguments that
one uses to derive the spectral properties  of the Luttinger-Sy
one-particle Hamiltonian. Since our aim is to study the thermodynamic properties and
Bose-Einstein condensation in this model, it is useful to derive the integrated
density of states first for a \textit{finite} system.\\
\textbf{3.1} Let $\Lambda:=[-L/2,L/2] \subset \mathbb{R}$. Then the
\textit{finite} Luttinger-Sy model with the Dirichlet boundary
conditions at $x=\pm L/2$ and with $n-1$ singular point-repulsive
($a\rightarrow +\infty$) impurities corresponds to the
one-dimensional self-adjoint Schr\"{o}dinger operator
\begin{equation}\label{hamil-L/2}
h_{L,X_n}:= \bigoplus_{j=1}^{n} h_{D}(I_{j})\ ,
\end{equation}
{acting in the direct orthogonal sum} of  Hilbert spaces (\ref{space-orth-sum}):
\begin{equation}\label{Hilb-spase-L/2}
\mathcal{H}_{\Lambda} := \bigoplus_{j=1}^{n} \mathcal{H}_{j} \ .
\end{equation}
Here
\begin{equation}\label{X_n}
X_n = \left\{x_0=-L/2 < x_1 < x_2 < \ldots < x_{n-1} <
x_{n}=L/2\right\} \,\,,\,\, \left\{I_j =
(x_{j-1},x_j)\right\}_{j=1}^{n} \,,
\end{equation}
and operators $\left\{h_{D}(I_{j})\right\}_{j=1}^{n}$ are defined
by (\ref{ham-Dirich}).

To make this system \textit{disordered}{,} Luttinger and Sy
\cite{Lutt-Sy-1} supposed that {the} impurity positions are random
variables, which are \textit{independently} and \textit{uniformly}
distributed over the interval $\Lambda$. Then instead of (\ref{X_n}) one
gets the random sets $\left\{X_{n}^\omega\right\}_
{\omega\in\Omega}$ on $(\Omega,\mathcal{F},\mathbb{P})$, which
a.s. contain $n-1$ ordered impurities
$\left\{x_j^{\omega}\right\}_{j=1}^{n-1}$. We denote the
corresponding random Luttinger-Sy Hamiltonian and eigenfunctions in $\Lambda$  by
\begin{equation}\label{random-hamil-L/2}
h_{D,n,L}^{\omega}:=h_{L,X_{n}^{\omega}} = \bigoplus_{j=1}^{n}
h_{D}(I_{j}^\omega) \ , \ \ h_{D,n,L}^{\omega}\Psi_{s_j,D,n}^{L,\omega} =
E_{s_j}(L_{j}^\omega)\Psi_{s_j,D,n}^{L,\omega} \ ,
\end{equation}
where the eigenfunctions have the form:
\begin{equation}\label{eigenfunct-L/2}
\Psi_{s_j,D,n}^{L,\omega}= (0,0,\ldots,\psi_{j,s_j}^{\omega}(x),\ldots,0)
\in  \bigoplus_{j=1}^{n} \mathcal{H}_{j}\ ,
\end{equation}
see definitions (\ref{ham-orthg-sum})-(\ref{eigenvalues-orth-space}).

Notice that Remark \ref{spectral-L-S-1} is valid in this case
modulo the substitution of $h_{D}^{\omega}$ by the random operator
(\ref{random-hamil-L/2}). In particular, for the spectrum of
(\ref{random-hamil-L/2}) one gets representation:
\begin{equation}\label{spect-L-S-n}
\sigma(h_{D,n,L}^{\omega})= \bigcup_{j=1}^{n}
\sigma(h_D (I_{j}^{\omega})).
\end{equation}

The following proposition is an immediate consequence of the
hypothesis about the independent uniform impurities distribution and the
\textit{thermodynamic limit}: $L\rightarrow\infty , n \rightarrow
\infty$, with a fixed density of impurities
\begin{equation}\label{lambda}
\lambda = \lim_{L\rightarrow\infty} \ \frac{n}{L} \ .
\end{equation}
\begin{proposition}\label{distributions}
\emph{(a)} In the thermodynamic limit the above finite-volume
random point field  $\left\{X_{n}^\omega\right\}$ converges
(\textit{in distribution}) to the Poisson point field
$\left\{X^\omega\right\}$ with intensity $\lambda$ and
corresponding random Poisson measure (\ref{Ran-Poiss-mes}).\\
\emph{(b)} The uniform and independent distribution of $n-1$
points of impurities induces on $\Lambda$ a random sets of intervals
$\left\{I_{j}^{\omega}\right\}_{j=1}^{n}, \omega\in\Omega$, of
random lengths $\left\{L_{j}=L_{j}^{\omega}\right\}_{j=1}^{n}$. The corresponding
joint probability distribution is
\begin{eqnarray}\label{n-distribution}
\label{measure-L} dP_{L,n}(L_1,...,L_n) = \frac{(n-1)!}{L^{n-1}} \ \delta(L_1+...+L_n - L) \
\ dL_1 dL_2 \ldots dL_n \ .
\end{eqnarray}
\emph{(c)} In the thermodynamic limit the lengths
$\left\{L_{j}^{\omega}\right\}_j$ form an infinite set of
independent random variables and distribution corresponding to
(\ref{n-distribution})  converges (weakly) to the
product-measure distribution ${\sigma}_{\lambda}$ defined by the set of consistent marginals:
\begin{eqnarray}\label{marginals}
d{\sigma}_{\lambda, k}(L_{j_1},\ldots,L_{j_k})=\lambda^k\prod_{s=1}^{k}e^{-\lambda
L_{j_s}}dL_{j_s} \ .
\end{eqnarray}
\end{proposition}
The proof is standard \cite{Bi,Sh}, see e.g. \cite{Le} for details.

Recall  that the finite-volume \textit{integrated density of
states} is defined by specific \textit{counting-function} \cite{PastFig}. For operator
(\ref{random-hamil-L/2}), it is a \textit{random} variable of the form:
\begin{equation}\label{fin-vol-IDS}
\mathcal{N}^\omega_L(E):=
\frac{1}{L}\sum_{\left\{\Psi_{s,D,n}^{L,\omega}\right\}}\theta(E
- E_{s,D}^{\omega}(n,L)) = \frac{1}{L}\int_{-L/2}^{L/2} dx \,\, \theta(E -
h_{D,n,L}^{\omega})(x,x) \ .
\end{equation}
Here $\theta(E - h_{D,n,L}^{\omega})(x,y)$ is the kernel of the
spectral-projection operator of $h_{D,n,L}^{\omega}$ corresponding
to the half-line $(-\infty, E)$ and $\theta(x)= \mathbb{I}_{(0, +\infty)}(x)$ stands for
the \textit{step-function}.
\begin{proposition}\label{L-S-IDS}
In thermodynamic limit the finite-volume integrated density of
states (\ref{fin-vol-IDS}) converges a.s. to non-random function
\begin{eqnarray}
\mathcal{N}_{\lambda}(E): =\lambda  \ \frac{e^{-c \lambda /\sqrt{E}}}{1-e^{-c \lambda
/\sqrt{E}}} \ , \label{IDOS-lim}
\end{eqnarray}
with $c=\pi/\sqrt{2}$.
\end{proposition}
\textit{Proof}: Explicit expressions (\ref{eigenvalues})
and (\ref{eigenfunctions}) imply for
(\ref{fin-vol-IDS}) the representation:
\begin{equation}\label{fin-vol-IDS-repr}
\mathcal{N}^\omega_L(E)=
\frac{1}{L}\sum_{j=1}^{n}\sum_{s=1}^\infty\theta\left\{E-\left(\frac{c
s}{L_{j}^\omega}\right )^2\right\} \ .
\end{equation}
Then by Proposition \ref{distributions} and by (\ref{lambda}), (\ref{fin-vol-IDS-repr}) we obtain
\begin{eqnarray}
&&{\mathcal{N}}_{\lambda}(E):= a.s.\lim_{L\to\infty}{\mathcal{N}}^\omega_{L}(E)=
a.s.\lim_{n\to\infty} \ \frac{\lambda}{n} \ \sum_{j=1}^{n}\sum_{s=1}^\infty\theta\left\{E-\left(\frac{c
s}{L_{j}^\omega}\right )^2\right\} =  \nonumber \\
&&\lambda \ \mathbb{E}_{\sigma_\lambda}\left\{ \sum_{s=1}^
\infty\theta\left(E-\left(\frac{c s}{L_i}\right )^2\right)\right\}
=\lambda^2\sum_{s=1}^\infty\int_0^\infty dL_i e^{-\lambda
L_i}\theta\left(E-\left(\frac{c s}{L_i}\right )^2\right) \ . \label{lim-IDS-repr-0}
\end{eqnarray}
The {a.s. limit} for (non-random) integrated density of states
$\mathcal{N}_{\lambda}(E)$ exists
by the \textit{Birkhoff ergodic theorem} \cite{Bi,Sh} and the uniform convergence of the $s$ - sum
ensures the permutation of expectation with respect the $\sigma_\lambda$ - distribution
(\ref{marginals}) and the sum. Thus, we obtain:
\begin{eqnarray}\label{lim-IDS-repr}
\mathcal{N}_{\lambda}(E)&=&\lambda^2\sum_{s=1}^\infty\int_{{c
s}/{\sqrt{E}}}^\infty dL_i e^{-\lambda
L_i}=\lambda\sum_{s=1}^\infty e^{-c s \lambda/\sqrt{E}} \ ,
\end{eqnarray}
which yields the explicit formula (\ref{IDOS-lim}). \qquad \hfill $\square$\\
\textbf{3.2} Formula (\ref{IDOS-lim}) allows us to recover for all energies $E>0$ the
one-dimensional integrated density of states for the free operator $t$, i.e. the case when density of impurities
$\lambda = 0$, cf. Remark \ref{propert-Poisson-example-1} and (\ref{Schr2}):
\begin{eqnarray}\label{zero-density}
\lim_{\lambda\downarrow0}\mathcal{N}_{\lambda}(E)= \mathcal{N}_{\lambda=0}(E)=
\frac{\sqrt{2}}{\pi}\sqrt{E} \ .
\end{eqnarray}
Notice that for the \textit{Lebesgue-derivative} $n_{\lambda}(E):=d\mathcal{N}_{\lambda}(E)/dE$,
i.e. for the \textit{density of states}  \cite{LMW} Sect.4, this limit  is not uniform in $E$ in the vicinity of the
spectrum edge $E=0$. This confirms the argument, previously presented in \cite{LPZ}, that the
Bose-Einstein condensation in such random media can not be {viewed} as a
perturbation of the free case, since this phenomenon is tightly
related to the behaviour of $n_{\lambda}(E)$ near the edge \cite{VDB-Lew-Pu, LPZ}.

On the other hand, {for $\lambda > 0$ and for $E$ close to the edge of the spectrum,}
the integrated density of states (\ref{IDOS-lim}) exhibits  the \textit{Lifshitz' tail} behaviour:
\begin{eqnarray}\label{Lif-tail}
\mathcal{N}_{\lambda}(E)=\lambda e^{- c \lambda /\sqrt{E}} \ \{1-O(e^{-2 c \lambda /\sqrt{E}})\} \ ,
\ E\downarrow 0 \ ,
\end{eqnarray}
see Remark \ref{propert-Poisson-example-1}. In this case $\lim_{E \downarrow 0} n_{\lambda}(E) = 0$.

It is known \cite{Kotani, LGP} that behaviour (\ref{Lif-tail}) near the edge remains valid even if
the parameter $ a>0 $ in (\ref{ran-delta1}) is \textit{finite}.
Notice that this parameter does not appear in the
leading term of the asymptotics (\ref{Lif-tail}). This can be explained by the fact that particle
with small energy  "sees" a point impurity potential with relative amplitude
$a/E \gg  1$. Therefore, in spite of its singular nature the Luttinger-Sy Hamiltonian seems to be a
good approximation for studying the BEC in Poisson random systems with non-singular repulsive impurities.

\section{Thermodynamics and Bose-Einstein Condensation}
\setcounter{equation}{0}
\renewcommand{\theequation}{\arabic{section}.\arabic{equation}} 

The second quantization of the one-particle Luttinger-Sy Hamiltonian (\ref{random-hamil-L/2}) in
the boson Fock space gives the one-dimensional PBG {embedded into a random potential}
created by
Poisson repulsive impurities (\ref{ran-delta1}) with $a = + \infty$. The latter implies that
bosons are distributed over independent intervals ("boxes")  $\left\{L_{j}^\omega \right\}_{j , \omega}$.\\
\textbf{4.1} We study the boson Luttinger-Sy model in the grand canonical ensemble,
defined by the inverse temperature $\beta>0$ and the chemical
potential $\mu$. Since the model corresponds to independent "boxes"
$\left\{L_{j}^\omega \right\}_{j , \omega}$,
the grand partition function of the PBG in $\Lambda =[-L/2,L/2]$ is the product of partition functions
calculated in individual "boxes" :
\begin{eqnarray*}
\Xi_L^{\omega}(\beta,\mu)=\prod_{j=1}^n \Xi_{L_j}^{\omega}(\beta,\mu) =
\prod_{j=1}^n \prod_{s=1}^\infty\left(1-e^{-\beta(E_{s_j}(L_{j}^\omega)-\mu)}\right)^{-1} \ ,
\end{eqnarray*}
see (\ref{random-hamil-L/2}). This gives for the grand canonical pressure
\begin{eqnarray} \label{pressure-2}
p_L^{\omega}(\beta,\mu)= -\frac{1}{\beta
L}\sum_{j=1}^n \sum_{s=1}^\infty\ln\left(1-e^{-\beta(E_{s_j}(L_{j}^\omega)-\mu)}\right) \ .
\end{eqnarray}
To ensure the convergence in (\ref{pressure-2}) we have to bound
chemical potential from above:  $\mu < \inf_{s_j, \omega}E_{s_j}(L_{j}^\omega)$.
By virtue of (\ref{eigenvalues}) we obtain in the thermodynamic limit:
\begin{equation}\label{mu-bound}
a.s.\lim_{L\rightarrow \infty}\inf_{s_j, \omega}E_{s_j}(L_{j}^\omega)= 0 \ .
\end{equation}
\blem For $\mu<0$ and $L\rightarrow \infty$ the pressure $p_L^{\omega}(\beta,\mu)$
converges almost surely to the non-random function
\begin{eqnarray}\label{pressure-lim}
p(\beta,\mu)= a.s.\lim_{L\to\infty}p_L^{\omega}(\beta,\mu)=-\frac{1}{\beta}\int_0^\infty
dE \ n_{\lambda}(E)\ln\left(1-e^{-\beta(E-\mu)}\right) \ ,
\end{eqnarray}
where the limiting density of states
\begin{eqnarray}
n_{\lambda}(E):=\frac{\lambda^2c}{2}\frac{e^{c \lambda /\sqrt{E}}}{E^{3/2}\left(e^{c \lambda /\sqrt{E}}-
1\right)^2} \ \ , \label{DOS}
\end{eqnarray}
and $c=\pi/\sqrt{2}$, cf. (\ref{IDOS-lim}).
\elem
\textit{Proof}: By definition of integrated density of states (\ref{fin-vol-IDS-repr}) we
can represent the pressure in (\ref{pressure-2}) as the Lebesgue-Stieltjes integral
\begin{equation*}
p_L^{\omega}(\beta,\mu)= -\frac{1}{\beta}\int_{0}^{\infty}\ d\mathcal{N}^\omega_L(E)
\ln\left(1-e^{-\beta(E-\mu)}\right) \ .
\end{equation*}
Then by virtue of (\ref{lim-IDS-repr-0}) we obtain that the limit
\begin{eqnarray*}
p(\beta,\mu)&=&a.s.\lim_{L\to\infty}p_L^{\omega}(\beta,\mu) \\
&=&-\frac{\lambda^2}{\beta}\sum_{s=1}^{\infty}\int_0^\infty dL_i
e^{-\lambda L_i}\ln\left(1-e^{-\beta((c s/L_i)^2-\mu)}\right) \
\end{eqnarray*}
exists a.s. and{, after change of variables and calculation of the sum,} takes the form:
\begin{eqnarray*}
p(\beta,\mu)=-\frac{\lambda^2}{\beta} \int_0^\infty
\frac{dk}{k^2}\frac{e^{-c \lambda /k}}{1-e^{-c \lambda/k}}\ln\left(1-e^{-\beta(k^2-\mu)}\right) \ .
\end{eqnarray*}
Setting $k=\sqrt{E}$, we recover the relation (\ref{pressure-lim}) with density of states (\ref{DOS}).
\qquad \hfill $\square$
\smallskip

Similarly we obtain the statement about the thermodynamic limit of
the grand-canonical particle density.
\blem \label{particle-density}
For all $\mu<0$ and $\beta>0$, the grand-canonical particles density
\begin{eqnarray}\label{density-L}
\rho^{\omega}_{L}(\beta,\mu)=\frac{1}{L} \
\sum_{j=1}^n\sum_{s=1}^\infty\frac{1}
{e^{\beta(E_{s_j}(L_{j}^\omega)-\mu)}-1} =
\int_{0}^{\infty}\ d\mathcal{N}^\omega_L(E)\ \frac{1}{e^{\beta(E - \mu)}-1} \ ,
\end{eqnarray}
converges a.s. to
\begin{eqnarray}\label{density-lim}
\rho(\beta,\mu)=\int_0^\infty dE \  \frac{n_{\lambda}(E)}{e^{\beta(E-\mu)}-1} \ ,
\end{eqnarray}
with density of states $n_{\lambda}(E)$ defined by (\ref{DOS}).
\elem
\textit{Proof}: By virtue of representation (\ref{density-L}), {the}
demonstration follows the same line of reasoning as we used above for the limiting pressure
(\ref{pressure-lim}). \qquad \hfill $\square$
\begin{corollary}\label{Crit-Dens}
By explicit formula (\ref{DOS}) we obtain that for the Luttinger-Sy model, defined by the Hamiltonian
(\ref{ham-orthg-sum}), the critical density
\begin{eqnarray}\label{critical-density}
\rho_c(\beta)=\lim_{\mu\uparrow0}\int_0^\infty
dE\frac{n_{\lambda}(E)}{e^{\beta(E-\mu)}-1} \
\end{eqnarray}
is bounded.
\end{corollary}
\textbf{4.2} It is known that for PBG the condition $\rho_c(\beta)< \infty$ implies the existence
of (\textit{generalized})
Bose condensation \cite{VDB-Lew-Pu}, when the particles density $\rho>\rho_c(\beta)$.
To make it obvious in our case we have to study  solutions
$\mu_L^\omega(\beta,\rho)$ of the finite-volume equations,
see (\ref{density-L}):
\begin{eqnarray}
\rho=\rho_L^\omega(\beta,\mu) \ , \ \omega \in \Omega \ .
\label{ran-mu-ro1}
\end{eqnarray}
In fact, the asymptotic behaviour of $\mu_L^\omega(\beta,\rho)$
studied for a general ergodic non-negative random potential in \cite{LPZ}.
These results then can be applied to the Luttinger-Sy model and lead to the following proposition:
\bprop \label{rand-BEC}
Let $\mu_L^\omega(\beta,\rho)$ be solution of the
equation (\ref{ran-mu-ro1}) for a given $\omega \in \Omega$. Then
\newline
{\rm{(a)}} for $\rho <\rho _{c}(\beta )$  the limit
\begin{equation}
a.s.\lim_{L\rightarrow \infty }\mu _{L}^{\omega }(\beta ,\rho
)=\mu (\beta ,\rho ) < 0  \ , \label{ran-mu-ro-lim1}
\end{equation}
exists and is the unique root of equation defined by (\ref{density-lim}):
\begin{equation}
\rho =\rho (\beta ,\mu ) \ ,  \label{mu-ro2}
\end{equation}
{\rm{(b)}} for $\rho \geq \rho _{c}(\beta )$ the limit
\begin{equation}
a.s.\lim_{L\rightarrow \infty }\mu _{L}^{\omega }(\beta ,\rho
)=0 \ , \label{ran-mu-ro-lim2}
\end{equation}
\eprop \

For $\rho \geq \rho _{c}(\beta )$  the limit (\ref{ran-mu-ro-lim2}) implies that the
density of condensed particles can be define in the usual (for \textit{generalized} condensation)
way:
\begin{equation}
\rho _{0}(\beta ,\rho ):=\lim_{\epsilon \downarrow
0}\,\left\{a.s. \ \lim_{L\rightarrow \infty }\int_{0}^{\epsilon
}\mathcal{N} _{L}^{\omega }(dE)\frac{1}{e^{\beta (E-\mu
_{L}^{\omega }(\beta ,\rho
))}-1}\right\}=\rho-\rho_c(\beta) \ , \label{ran-BEC-lim}
\end{equation}
see e.g. \cite{VDB-Lew-Pu}.
If $\rho <\rho _{c}(\beta )$, the limit is zero. Notice that this nonrandom limit is a consequence
of the chemical potential asymptotics (Proposition \ref{rand-BEC}) and of the uniform
convergence of the particle density (\ref{density-L}), see \cite{LPZ}, Theorem 4.1.

\section{Off-Diagonal Long-Range Order}
\setcounter{equation}{0}
\renewcommand{\theequation}{\arabic{section}.\arabic{equation}} 

In this section we study the problem the \textit{two-point} correlation function \cite{OP,Yang}.
By definition, in the finite volume $\Lambda$ and for any $\omega \in \Omega$, it has the form:
\begin{eqnarray}
\rho^{L}_\omega (x,y; \beta ,\mu):&=&\sum_{j=1}^n \sum_{s_j=1}^\infty\frac{1}{e^{\beta
(E_{s_j}(L_{j}^\omega)-\mu )}-1 } \
\left(\overline{\Psi_{s_j,D,n}^{L,\omega}(x)}, \Psi_{s_j,D,n}^{L,\omega}(y)\right)_{\mathbb{R}^n}
\nonumber \\
&=& \sum_{j=1}^n \sum_{s_j=1}^\infty\frac{1}{e^{\beta
(E_{s_j}(L_{j}^\omega)-\mu )}-1 }\overline{\psi
_{j,s}^{\omega}}(x) \ \psi _{j,s}^{\omega}(y) \ , \label{red-dens-matr}
\end{eqnarray}
where $(\cdot \,,\, \cdot)_{\mathbb{R}^n}$ denotes the scalar product in $\mathbb{R}^n$,  see
(\ref{random-hamil-L/2}) and (\ref{eigenfunct-L/2}). Therefore (\ref{red-dens-matr})
is the kernel of for the \textit{one-body reduced density matrix}, see e.g. \cite{LewPul}.\\
\textbf{5.1} We know that this function is \textit{not
self-averaging} in the thermodynamic limit \cite{LGP,
PastFig}. {To get a way out,} we proposed in \cite{LPZ} to consider the
\textit{space-averaged} version of (\ref{red-dens-matr}):
\begin{equation}
\tilde{\rho}^{L}_\omega (x,y; \beta ,\mu ):=\frac{1}{L}
\int_{-L/2}^{L/2}dz \ \rho^{L}_\omega(\beta ,\mu ;x+z,y+z) \ .
\label{aver-rand-dens-matr-1}
\end{equation}
The motivation was based on the fact that in the limit $\lambda \downarrow 0$, we
recover from (\ref{aver-rand-dens-matr-1}) the free case, see \cite{LPZ}
and Section 5.2 below.
\btheo For the Luttinger-Sy model, the thermodynamic limit of the {space-averaged} two-point
correlation function (\ref{aver-rand-dens-matr-1}) for $\beta>0$
and $\mu\leq0$, is a.s. nonrandom and has the form:
\begin{eqnarray}\label{corr-func-lim}
\tilde{\rho}(x,y;\beta,\mu)=\rho_0(\beta,\rho)+e^{-\lambda|x-y|}\int_0^\infty
dE\frac{n_{\lambda}(E)}{e^{\beta(E-\mu)}-1}\cos(\sqrt{2E}(x-y)) \ .
\end{eqnarray}
Here  $n_{\lambda}(E)$ is defined by (\ref{DOS}) and $\rho_0(\beta,\rho)$ the
condensate density  (\ref{ran-BEC-lim}).
\etheo\
\textit{Proof}: We consider first the case of negative chemical potential (\ref{ran-mu-ro-lim1}),
i.e. $\rho_0(\beta,\rho)= 0$. Using explicit form of eigenfunctions (\ref{eigenfunctions},
we obtain for the thermodynamic limit of (\ref{aver-rand-dens-matr-1}):
\begin{eqnarray}\label{ODLRO-1}
&&\tilde{\rho}(x,y;\beta,\mu)=\lim_{L\to\infty}\frac{\lambda}{n} \ \sum_{j=1}^n \sum_{s=1}^\infty
\frac{1}{e^{\beta(E_{s_j}(L_{j}^\omega)-\mu)}-1}\times \\
&&\frac{2}{L_j}\int_0^Lda\sin(k_{s_j}(L_{j}^\omega)(x+a-y_{j-1}^\omega))
\sin(k_{s_j}(L_{j}^\omega)(y+a-y_{j-1}^\omega))\mathbb{I}_{I_{j}^{\omega}}(x+a)
\mathbb{I}_{I_{j}^{\omega}}(y+a) \ ,\nonumber
\end{eqnarray}
with $k_{s_j}(L_{j}^\omega):=\sqrt{2E_{s_j}(L_{j}^\omega)}$. Let us put, for simplicity, $x>y$.
Then the integration is reduced to $[y_j^\omega-y,y_j^\omega-x+L_j^\omega]$, such that
$(x-y)\leq L_j^\omega$. Since $k_{s_j}(L_{j}^\omega) L_j ^\omega=s\pi$, the integration  over $a$ yields
\begin{eqnarray*}
&&\frac{2}{L_j}\int_0^L da\sin(k_{s_j}(L_{j}^\omega)(x+a-y_{j-1}^\omega))
\sin(k_{s_j}(L_{j}^\omega)(y+a-y_{j-1}^\omega))
\mathbb{I}_{I_{j}^{\omega}}(x+a)\mathbb{I}_{I_{j}^{\omega}}(y+a)=
\nonumber\\
&&\cos(k_{s_j}(L_{j}^\omega)(x-y))\theta(L_j^\omega-(x-y))\left(1-\frac{x-y}{L_j^\omega}\right),
\end{eqnarray*}
with step function $\theta(z)$. Since by Proposition \ref{distributions}
random variables $L_j^\omega$ are independent in the limit $L\rightarrow\infty$, we apply to
(\ref{ODLRO-1}) the Birkhoff ergodic theorem and find the limit:
\begin{eqnarray*}\label{correl-func-1}
\tilde{\rho}(x,y;\beta,\mu)&=& a.s. \ \lambda^2 \ \sum_{s=1}^\infty\int_0^\infty
dL_j e^{-\lambda L_j} \frac{1}{e^{\beta((c s/L_j)^2)
-\mu)}-1}\times\nonumber\\&&
\cos(\sqrt{2}c s(x-y)/L_j)\theta(L_j-(x-y))\left(1-\frac{x-y}{L_j}\right)\ .
\end{eqnarray*}
with $c=\pi/\sqrt{2}$. If we put $q = cs/L_j$, then
\begin{eqnarray}\label{ODLRO-2}
\tilde{\rho}(x,y;\beta,\mu)&=&\lambda^2 \ \sum_{s=1}^\infty\int_0^\infty
\frac{dq}{q^2} \ e^{- cs \lambda/q} \
\frac{1}{e^{\beta(q^2-\mu)}-1}\times \\&&
\cos(\sqrt{2}q(x-y)) \ \theta(s - q(x-y)/c) \ c \ \{s- q(x-y)/c\} \ . \nonumber
\end{eqnarray}
The sum over $s$  yields
\begin{eqnarray}
\sum_{s \geq s_{min}}^{\infty} e^{- cs \lambda /q}(s- q(x-y)/c)=e^{-\lambda(x-y)}\frac{e^{-c \lambda
/q}}{(1-e^{-c\lambda /q})^2} \ ,
\end{eqnarray}
where $s_{min}= [q(x-y)/c]$ denotes the \textit{entire} part of $ q(x-y)/c$. Then after change of
variables,
$q=\sqrt{E}$, we find by (\ref{DOS}) and (\ref{ODLRO-2}) for $\mu<0$:
\begin{eqnarray}\label{correl-func-2}
\tilde{\rho}(x,y;\beta,\mu)= \ e^{-\lambda |x-y|} \ \int_0^\infty
dE \ \frac{n_{\lambda}(E)}{e^{\beta(E-\mu)}-1} \ \cos(\sqrt{2E}(x-y)) \ .
\end{eqnarray}
We put here $|x-y|$, since the proof for $x-y \leq 0$ is identical to that for $0 \leq x-y$.

Now we shall study the case when the condensate exists. The finite-volume chemical potential
$\mu_L^\omega(\beta,\rho)$ is a solution of equation (\ref{ran-mu-ro1}). By
Proposition \ref{rand-BEC}(b)
for $\rho>\rho_c(\beta)$ it implies (\ref{ran-mu-ro-lim2}), i.e.
$\mu_L^\omega(\beta,\rho>\rho_c(\beta))$
converges a.s. to $0$. To find the limit of {space-averaged} correlation function
(\ref{aver-rand-dens-matr-1})
for the sequence $\left\{\mu_L^\omega(\beta,\rho)\right\}_{L}$  we split (\ref{aver-rand-dens-matr-1})
into two parts:
\begin{eqnarray}\label{corr-func-cond}
&&\tilde{\rho}_L^\omega(x,y;\beta,\mu_L^\omega(\beta,\rho))=
\frac{1}{L} \ \sum_{j=1}^n \sum_{s=1}^\infty\frac{1}
{e^{\beta(E_{s_j}(L_{j}^\omega)-\mu_L^{\omega}(\beta,\rho))}-1} \times \\
&&\cos\left(\sqrt{E_{s_j}(L_{j}^\omega)}(x-y)\right)
\theta(L_j^\omega-(x-y))\left(1-\frac{(x-y)}{L_j^\omega}\right)
\theta(\delta-E_{s_j}(L_{j}^\omega))\nonumber\\
&&+ \ \frac{1}{L} \ \sum_{j=1}^n \sum_{s=1}^\infty\frac{1}
{e^{\beta(E_{s_j}(L_{j}^\omega)-\mu_L^{\omega}(\beta,\rho))}-1} \times \nonumber\\
&&\cos\left(\sqrt{E_{s_j}(L_{j}^\omega)}(x-y)\right)\theta(L_j^\omega-(x-y))
\left(1-\frac{(x-y)}{L_j^\omega}\right)\theta(E_{s_j}(L_{j}^\omega)-\delta)\ , \nonumber
\end{eqnarray}
for some  $\delta>0$. Since in the second term of the right-hand side of (\ref{corr-func-cond}) the
eigenvalues $E_{s_j}(L_{j}^\omega)\geq \delta>0$, the limit (\ref{ran-mu-ro-lim2}) and uniform
convergence of the sums (cf. Corollary 3.1 in \cite{LPZ}) yields
\begin{eqnarray}
&&\lim_{\delta\downarrow0}\lim_{L\to\infty}\tilde{\rho}_L^\omega(x,y;\beta,
\mu_L^\omega(\beta,\rho))= a.s. \ \lim_{\delta\downarrow0}\lim_{L\to\infty}\frac{1}{L}
\sum_{i=1}^\nu\sum_{s=1}^\infty\frac{1}{e^{\beta(E_{s_j}(L_{j}^\omega)-
\mu_L^{\omega}(\beta,\rho))}-1} \times \nonumber\\
&&\cos\left(\sqrt{E_{s_j}(L_{j}^\omega)}(x-y)\right)\theta(L_i^\omega-(x-y))
\left(1-\frac{(x-y)}{L_i^\omega}\right)\theta(\delta-E_{s_j}(L_{j}^\omega))\nonumber\\
&&+ \ \tilde{\rho}(x,y;\beta,0) \ , \label{corr-func-cond-bis}
\end{eqnarray}
where the last term is defined by (\ref{correl-func-2}).

To study the limit of the first term in the right-hand side of
(\ref{corr-func-cond-bis}) we use the fact that the levels with
energies $E_{s_j}(L_{j}^\omega)<\delta$ are, by definition, of order
$O(\delta)$. {By (\ref{eigenvalues})} these levels are defined in
large "boxes" of lengths of order $O(\delta^{-1/2})$. Then for
$E_{s_j}(L_{j}^\omega)<\delta$ and $|x-y|\ll\delta^{-1/2}$ with
$\delta$ small enough, we obtain asymptotics
\begin{equation*}
\cos\left(\sqrt{E_{s_j}(L_{j}^\omega)}(x-y)\right)\theta(L_i^\omega-(x-y))
\left(1-\frac{(x-y)}{L_i^\omega}\right)=1+O\left(\sqrt{\delta}\right) \ .
\end{equation*}
Therefore, by definition of \textit{generalized} condensation (\ref{ran-BEC-lim})
and by (\ref{corr-func-cond-bis}),
we get
\begin{eqnarray}
\tilde{\rho}_L^\omega(x,y; \beta, 0):=
a.s. \ \lim_{\delta\downarrow 0}\lim_{L\to\infty}\tilde{\rho}_L^\omega(x,y;\beta,
\mu_L^\omega(\beta,\rho)) =
\rho_0(\beta,\rho)+\tilde{\rho}(x,y;\beta,0) \ ,
\end{eqnarray}
for $\rho>\rho_c(\beta)$. This finishes the proof of (\ref{corr-func-lim}). \qquad \hfill $\square$\\
\textbf{5.2} It is instructive to make a contact between the concept
of the \textit{space-averaged} one-body reduced density matrix
(\ref{aver-rand-dens-matr-1}) and the one for nonrandom free PBG.
\bcorol When density of particles $\rho$ exceeds the critical value
$\rho_c(\beta)$ (\ref{critical-density}), the
\textit{space-averaged} one-body reduced density matrix of the
Luttinger-Sy model, manifests Off-Diagonal Long-Range Order:
\begin{eqnarray}\label{ODLRO}
ODLRO(\beta,\rho):=\lim_{|x-y|\to\infty}\tilde{\rho}(x,y;\beta,\mu(\beta,\rho))
=\rho_0(\beta,\rho) \ ,
\end{eqnarray}
We see that similarly to the nonrandom case this limit is defined by the
condensation density (\ref{ran-BEC-lim}).
\ecorol
\begin{remark}\label{ODLRO-nonrand-lim}
Notice that for $\mu<0$ (or $\rho < \rho_c(\beta)$) , the space-averaged reduced density matrix
(\ref{corr-func-lim}) remains consistent with the free nonrandom case. Indeed, by (\ref{DOS}) we get
for (\ref{correl-func-2}) the limit
\begin{eqnarray}
\lim_{\lambda\downarrow0}\tilde{\rho}(x,y;\beta,\mu)=\rho(x,y;\beta,\mu) \ ,
\end{eqnarray}
where
\begin{equation}\label{ODLRO-nonrand}
\rho(x,y;\beta,\mu)= \frac{1}{\pi}\int_0^\infty
\frac{dE}{\sqrt{2E}} \ \frac{1}{e^{\beta(E-\mu)}-1} \ \cos(\sqrt{2E}(x-y))
\end{equation}
coincides with two-point correlation function of the free PBG \cite{LPZ}.
This equivalence is valid only when there is no condensation, since by (\ref{DOS}) and
(\ref{critical-density}) one has that $\lim_{\lambda\downarrow0}\rho_c(\beta) = \infty$.
\end{remark}

Let the single-impurity potential $u(x)\geq 0,\,x\in \mathbb{R}$,  have  compact support
$[-\eta/2,\eta/2]$, see Section 2.2. Then one has
\bprop{\rm\textbf{\cite{LPZ}}} \label{1-dim-ODLRO-1}
Let $\tilde{\gamma}:=1-e^{-\tilde{v}}$, with
$\tilde{v}:=\int_{\mathbb{R}}v(x)dx$. Then
\begin{equation}  \label{1-dim-est-ODLRO}
\tilde{\rho}(x,y;\beta,\mu)\leq \rho(x,y;\beta,\mu) \ e^{-\lambda
\tilde{\gamma} (|x - y| - \eta)} \ ,
\end{equation}
$\mathbb{P}-$a.s. for any $\mu<0$.
\eprop
Since for the Luttinger-Sy model the single-impurity potential  is defined
as a singular $\delta$-point potential with infinite amplitude (Section 2.4), we get in
(\ref{1-dim-est-ODLRO}) $\tilde{\gamma}=1$ and $\eta = 0$.

To check this directly, put $E=w|x-y|^2$. Then for $\mu<0$ we can represent two-point correlation function
(\ref{correl-func-2}) as
\begin{eqnarray}\label{corr-func-lim-bis}
&&\tilde{\rho}(x,y;\beta,\mu)=  {Re}\ e^{-\lambda |x - y|}\ \sum_{s=1}^\infty
e^{\beta\mu s}\int_0^\infty dE e^{-\beta E s}n_{\lambda}(E)e^{i\sqrt{2E}|x-y|} \\
&&={e^{-\lambda |x - y|}}{|x-y|^2}\ {Re} \ \sum_{s=1}^\infty e^{\beta\mu
s}\int_0^\infty dw \ n_{\lambda}(w|x-y|^2)\ e^{-|x-y|^2(\beta w s-i\sqrt{2w})} \  \nonumber .
\end{eqnarray}
To calculate  the asymptotics of $\tilde{\rho}(x,y;\beta,\mu)$, when $|x-y|\rightarrow \infty$, we
estimate the last integral in (\ref{corr-func-lim-bis}) by the \textit{saddle-point} method. Then
(\ref{DOS}) implies
\begin{eqnarray*}
\tilde{\rho}(x,y;\beta,\mu) &=& e^{-\lambda |x - y|} \ \left\{\sum_{s=1}^\infty
e^{\beta\mu s}\ \frac{e^{-{|x-y|^2}/{4\beta s}}}{(2\pi\beta s)^{1/2}} +
e^{- \sqrt{2|\mu|} |x - y|} O (|x-y|^{-1})\right\} \nonumber \\
&=& e^{-\lambda |x - y|} \ \ \rho(x,y;\beta,\mu) \, + \,
e^{- (\sqrt{2|\mu|} + \lambda) |x - y|} \ \, O (|x-y|^{-1})  \ , \label{corr-func-asympt}
\end{eqnarray*}
for $|x-y|\rightarrow \infty$, with the free PBG two-point correlation function
$\rho(x,y;\beta,\mu)$ defined by (\ref{ODLRO-nonrand}).  This
confirms the statement (\ref{1-dim-est-ODLRO}) for the case of the Luttinger-Sy model.

\section{Comments and discussion}
\setcounter{equation}{0}
\renewcommand{\theequation}{\arabic{section}.\arabic{equation}} 

\textbf{6.1 Critical density}

\smallskip

\noindent We start by remark concerning modifications of the Luttinger-Sy model properties,
and in particular of the value of the critical density,
when one passes from infinite to a \textit{finite} amplitude $a < \infty$ of the $\delta$-potential
(\ref{ran-delta1}).

Recall that operators $\left\{h_{a}^{\omega}\right\}_{a\geq 0}$ correspond to a {monotonically}
increasing family of quadratic forms with $h_{a=+\infty}^{\omega}= h_{D}^{\omega}$, see (\ref{s.r.}).
Then by definition of the integrated density of states
(\ref{fin-vol-IDS}), (\ref{IDOS-lim}) and by the \textit{mini-max} principle for $h_{D}^{\omega}$ and
$h_{a}^{\omega}${,} one gets
\begin{equation}\label{IDS-ineq}
\mathcal{N}_{\lambda}(E) \equiv \mathcal{N}_{\lambda,a=+\infty}(E)< \mathcal{N}_{\lambda, a}(E) \leq
\mathcal{N}_{\lambda, a=0}(E)= \mathcal{N}_{\lambda=0, a}(E)=\frac{\sqrt{E}}{c}\ .
\end{equation}
Notice that the integrated density of states for the free case $a=0$ coincides with that for the zero
impurity density (\ref{zero-density}), $\lambda=0$. From (\ref{IDS-ineq}) one gets the corresponding
inequalities for critical densities (\ref{critical-density}) indicating the enhancement of the BEC,
that was remarked already by Kac and Luttiger \cite{KL1,KL2}.

More refined arguments (see e.g. \cite{PastFig}, Ch.III, 6B) give for $E < \pi^2 a^2/32$, the estimate,
cf. (\ref{lim-IDS-repr}):
\begin{equation}\label{IDS-estim}
\mathcal{N}_{\lambda}(E)< \mathcal{N}_{\lambda, a}(E)<
\lambda\sum_{s=1}^\infty e^{- s \lambda(c/\sqrt{E} - 4/a)}=:\mathcal{N}_{\lambda, a}^*(E) \ .
\end{equation}
Then, definition of the critical density
(\ref{critical-density}) and  (\ref{IDS-ineq}) yield
\begin{eqnarray}
\nonumber \rho_c(\beta,\lambda):&=&\lim_{\mu\uparrow0}\int_0^\infty
\frac{d\mathcal{N}_{\lambda}(E)}{e^{\beta(E-\mu)}-1} = \int_0^\infty \ dE \ \mathcal{N}_{\lambda}(E)
\ \frac{\beta e^{\beta E}}{(e^{\beta E}-1)^2}  \\
&\leq&\int_0^\infty \ dE \ \mathcal{N}_{\lambda,a}(E) \
\frac{\beta e^{\beta E}}{(e^{\beta E}-1)^2} =:\rho_c(\beta,\lambda,a) \ ,\label{critical-density-IDS}
\end{eqnarray}
with obvious limits: $\lim_{a\rightarrow 0} \rho_c(\beta,\lambda,a)= \infty$ and
$\lim_{\lambda\rightarrow 0} \rho_c(\beta,\lambda,a)= \infty$, by (\ref{IDS-ineq}).

The estimate (\ref{IDS-estim}) yields the upper bound on $\rho_c(\beta,\lambda,a)$ for small $a$.
Setting $\tilde{E}(a):= (\pi a /8)^2$ by (\ref{IDS-ineq})-(\ref{critical-density-IDS}) we get that
\begin{equation}\label{ineq-1}
\rho_c(\beta,\lambda,a)\leq \int_0^{\tilde{E}(a)} \ dE \ \mathcal{N}_{\lambda,a}^*(E) \
\frac{\beta e^{\beta E}}{(e^{\beta E}-1)^2} + \int_{\tilde{E}(a)}^\infty \ dE \ \frac{\sqrt{E}}{c} \
\frac{\beta e^{\beta E}}{(e^{\beta E}-1)^2} =: I (\beta,\lambda,a) \ .
\end{equation}
Then for fixed $\lambda > 0$ and small $a>0$ we obtain the estimate:
\begin{equation*}
\rho_c(\beta,\lambda,a) \leq
I (\beta,\lambda,a)\leq \frac{1}{\beta\lambda}\left(\frac{8}{\pi}\right)^2 \frac{1}{4e(\sqrt{2}-1)}
+ \frac{1}{a} \ \frac{16 \sqrt{2}}{\beta \pi^2}   \ .
\end{equation*}
Similarly, for $a > 0$ and small $\lambda>0$ we obtain by (\ref{ineq-1}) that
\begin{equation*}
\rho_c(\beta,\lambda,a) \leq
\frac{1}{\lambda \beta}\int_{0}^{\tilde{E}(a)}\ \frac{dx}{x^2} \frac{e^{-c/\sqrt{x}}}{1- e^{-c/\sqrt{x}}}
+ \int_{\tilde{E}(a)}^\infty \ dE \ \frac{\sqrt{E}}{c} \
\frac{\beta e^{\beta E}}{(e^{\beta E}-1)^2} \ .
\end{equation*}

Notice that the \textit{bounded} critical density $\rho_c(\beta,\lambda,a)$  for
$\lambda > 0$ and $a > 0$ is the key criterium of existence of BEC in the one-dimensional
system with integrated density of states $\mathcal{N}_{\lambda, a}(E)$, cf. (\ref{ran-BEC-lim}).
On the other hand the Bogoliubov-Hohenberg theorem says that there is no BEC in
\textit{translation invariant} boson systems if dimension is less than $d=2$, see e.g. \cite{Bog-2,
BunzMart}. Therefore, the BEC in the Luttinger-Sy model is of a different nature
than the case without random impurity potential.

In fact, the randomness of the impurity potential is not indispensable for BEC in
one-dimensional perfect Bose-gas. To this end we construct \textit{nonrandom hierarchical models}
with impurity potential which manifest the BEC via mechanism similar to that in the
Luttinger-Sy model.

\smallskip

\noindent \textbf{6.2 Hierarchical model for BEC in one-dimensional
nonrandom intervals}

\smallskip

\noindent We present here a nonrandom  \textit{hierarchical} one-dimensional
system, which manifests BEC  and in a certain sense
mimics the Luttinger-Sy model.

\smallskip

\noindent \textbf{Type I BEC.} Let $\Lambda:=(0,L)$ be a segment
separated into $n$ \textit{impenetrable} intervals of lengths $L_j$,
$j=1,...,n$ such that $\lambda=n/L<\infty$. For simplicity we take
the hierarchy when all intervals, except the first
(\textit{largest}) one, are identical:
\begin{equation}\label{L-ZL}
L_1=\frac{\ln(\lambda L)}{\lambda} \ \ \ \ {\rm{and}} \ \ \  \
L_{j\neq 1}={\tilde{L}_n}=\frac{L-L_1}{n-1} \ .
\end{equation}
Then one gets
\begin{equation}\label{}
\lim_{L\to\infty}L_1=+\infty \ \ \ \ {\rm{and}} \ \ \
\lim_{L\to\infty}{\tilde{L}_n}=\frac{1}{\lambda}  \ .
\end{equation}
This non-random system presents an obvious analogue of the
Luttinger-Sy model. Here again, the quantum states are defined in
independent intervals and they have energies
\begin{eqnarray}\label{energy-ZL}
E_{j,s}=\frac{c^2s^2}{L_j^2} \ , \ j=1,...,n \ , \ s=1,2,... \ ,
\end{eqnarray}
with $c^2=\pi^2/2$. The spectrum of the corresponding Schr\"{o}dinger
operator is discrete and bounded below by zero, cf.
(\ref{eigenvalues}), (\ref{spect-L-S}). Then the chemical potential
is $\mu<0$ and the PBG particle density in $\Lambda$ has the same
expression as in (\ref{density-L})
\begin{equation*}
\rho_L(\beta,\mu)=\frac{1}{L}\sum_{j=1}^{n}\sum_{s=1}^\infty\frac{1}
{e^{\beta(E_{j,s}-\mu)}-1} \ , \ \beta>0 \ , \ \mu \leq 0 \ .
\end{equation*}
By virtue of the hierarchical structure of intervals we can separate
the expression for density into two parts:
\begin{eqnarray}\label{dens-hierarch}
\rho_L(\beta,\mu)=\frac{1}{L}\sum_{s=1}^\infty\frac{1}{e^{\beta(c^2s^2/L_1^2-\mu)}-1}+
\frac{n-1}{L}\sum_{s=1}^\infty\frac{1}{e^{\beta(c^2s^2/{\tilde{L}_n}^2-\mu)}-1} \ .
\end{eqnarray}
Since $L_1=O(\ln(\lambda L))$,  the first sum in
(\ref{dens-hierarch}) converges, when $L\to\infty$, to zero for all $\mu \leq 0 $, i.e. we obtain
\begin{eqnarray}\label{ro-M=1}
\rho(\beta,\mu)=\lim_{L\to\infty}\rho_L(\beta,\mu)=\lambda\sum_{s=1}^\infty\frac{1}
{e^{\beta((cs/\lambda)^2-\mu)}-1} \ .
\end{eqnarray}
As a consequence, the critical density for this system is finite:
\begin{eqnarray}\label{ro-crit-M=1}
\rho_c(\beta):=\sup_{\mu \leq 0}\rho(\beta,\mu)=\rho(\beta,0)=\lambda\sum_{s=1}^\infty\frac{1}
{e^{\beta(cs/\lambda)^2}-1} < \infty \ ,
\end{eqnarray}
and we have BEC condensation, when $\rho>\rho_c(\beta)$.

This condensation is of \textit{type I}, since the difference
between the \textit{ground-state} energy and the energy of the
\textit{first excited} state (which are both localized in the
biggest interval $L_1$) is of the order $O(L_1^{-2})=O((\ln(\lambda
L))^{-2})$, see e.g. \cite{VDB-Lew-Pu}, or \cite{ZagBru}. In this
case the solution $\mu_L(\beta,\rho)$ of the equation
\begin{eqnarray}\label{dens-mu-eq}
\rho &=& \frac{1}{L}\frac{1}{e^{\beta(c^2/L_1^2-\mu)}-1}\ + \ \frac{1}{L}
\sum_{s>1}\frac{1}{e^{\beta((cs/L_1)^2-\mu)}-1}  \\ \nonumber
&+& \frac{\nu-1}{L}
\sum_{s=1}^\infty\frac{1}{e^{\beta((cs/{\tilde{L}_n})^2-\mu)}-1} \ ,
\end{eqnarray}
for $\rho>\rho_c(\beta)$ and large $L$ has asymptotics
\begin{equation}\label{mu-ZL}
\mu_L(\beta,\rho)=E_{1,1}-\frac{1}{\beta \rho_0(\beta, \rho)L}+O(1/L^2) \ ,
\end{equation}
see (\ref{energy-ZL}).
Inserting (\ref{mu-ZL}) into (\ref{dens-mu-eq}) we obtain in the limit
\begin{eqnarray*}
\rho&=&\lim_{L\to\infty} \ \frac{1}{L}\frac{1}{e^{\beta(c^2/L_1^2-\mu_L(\beta,\rho))}-1} \ + \
\lambda \ \sum_{s=1}^\infty\frac{1}{e^{\beta (cs/\lambda)^2}-1} \nonumber \\
&=&\rho_0(\beta, \rho)+\rho_c(\beta) \ ,
\end{eqnarray*}
where by (\ref{mu-ZL}) the \textit{condensate density} is
\begin{eqnarray*}
\rho_0(\beta, \rho)=\lim_{L\to\infty} \ \frac{1}{L} \ \frac{1}{e^{\beta(c^2/L_1^2-
\mu_L(\beta,\rho))}-1}  \ .
\end{eqnarray*}
Therefore, this one-dimensional hierarchical model shows {a}
\textit{type I} BEC localized in the (\textit{logarithmically})
large, but not macroscopic{, domain corresponding} to the
ground-state wave function. Generalizations to another hierarchy of
intervals with one largest interval trapping BEC are obvious.

For example, it is easy to generalize the above observation to the
\textit{type I} BEC localized in a \textit{finite} number of $M$
identical (\textit{logarithmically}) large intervals:
\begin{equation}\label{M}
L_1= \ldots =L_M=\frac{\ln(\lambda L)}{\lambda} \ \ \ \ {\rm{and}} \ \ \  \
L_{j}={\tilde{L}_n}=\frac{L- M L_1}{n-M} \ , \ M<j\leq n \ ,
\end{equation}
cf. (\ref{L-ZL}). Then similar to the case $M=1$ (\ref{dens-hierarch}) one gets
\begin{eqnarray}\label{ro-M}
\rho_L(\beta,\mu)=\frac{1}{L}\sum_{j=1}^M \
\sum_{s=1}^\infty\frac{1}{e^{\beta(c^2s^2/L_j^2-\mu)}-1}+
\frac{n-M}{L}\sum_{s=1}^\infty\frac{1}{e^{\beta(c^2s^2/{\tilde{L}_n}^2-\mu)}-1} \ .
\end{eqnarray}
which implies trough verbatim that for $M>1$ the critical density (\ref{ro-crit-M=1}) rests the
same. If $\rho>\rho_c(\beta)$, the equation $\rho = \rho_L(\beta,\mu)$ (\ref{ro-M})
yields for asymptotics of the solution $\mu_L(\beta,\rho)$ an expression similar
to (\ref{mu-ZL}) for $M=1$:
\begin{equation}\label{mu-ZL-M}
\mu_L(\beta,\rho)=E_{j,1}-\frac{M}{\beta\rho_0(\beta, \rho)L}+O(1/L^2) \ , \ 1 \leq j \leq M \ .
\end{equation}
Here $E_{1,1}=\ldots= E_{M,1}$ by (\ref{energy-ZL}) and (\ref{M}). Then taking into
account (\ref{energy-ZL}) and (\ref{M}), (\ref{mu-ZL-M}) we obtain by (\ref{ro-M})
\begin{eqnarray*}
\rho&=&\lim_{L\to\infty} \ \frac{1}{L}\sum_{j=1}^M \ \frac{1}{e^{\beta(c^2/L_j^2-\mu_L(\beta,\rho))}-1}
\ + \
\lambda \ \sum_{s=1}^\infty\frac{1}{e^{\beta (cs/\lambda)^2}-1} \nonumber \\
&=&\rho_0(\beta, \rho)+\rho_c(\beta) \ ,
\end{eqnarray*}
where the condensate density is \textit{equally} shared among the
first $M$ intervals:
\begin{eqnarray*}
\rho_0(\beta, \rho)=\lim_{L\to\infty} \ \frac{M}{L} \ \frac{1}{e^{\beta(c^2/L_1 ^2-
\mu_L(\beta,\rho))}-1} \ .
\end{eqnarray*}
\smallskip

\noindent \textbf{Type II BEC.} To obtain the \textit{type II} BEC
in \textit{one} interval we take, instead of (\ref{L-ZL}):
\begin{equation}\label{L-ZL-II}
L_1 := \sqrt{L /\lambda} \ \ \ \ {\rm{and}} \ \ \  \
L_{j\neq 1}={\tilde{L}_n}=\frac{L-L_1}{n-1} \ .
\end{equation}
Then (\ref{dens-hierarch}) gets the form
\begin{eqnarray}\label{dens-hierarch-II}
\rho_L(\beta,\mu)=\frac{1}{L}\sum_{s=1}^\infty \frac{1}{e^{\beta(\lambda c^2s^2/L - \mu)}-1}+
\frac{n-1}{L}\sum_{s=1}^\infty\frac{1}{e^{\beta(c^2s^2/{\tilde{L}_n}^2-\mu)}-1} \ .
\end{eqnarray}
Since for $\mu \leq 0$ the first sum in (\ref{dens-hierarch-II}) converges to zero, when
$L \rightarrow \infty$, we obtain for $\rho(\beta,\mu)=\lim_{L\to\infty}\rho_L(\beta,\mu)$
and $\rho_c(\beta)$ the same expressions as in (\ref{ro-M=1}) and (\ref{ro-crit-M=1}).

Now, if $\rho>\rho_c(\beta)$, the solution $\mu_L(\beta,\rho)$ of equation
$\rho=\rho_L(\beta,\mu)$ has asymptotics defined by (\ref{dens-hierarch-II}):
\begin{eqnarray}\label{dens-hierarch-II-lim}
\rho &=&
\lim_{L\to\infty} \ \frac{1}{L}\sum_{s=1}^\infty \frac{1}{e^{\beta(\lambda c^2s^2/L-\mu_L(\beta,\rho))}-1}
\ + \
\lambda \ \sum_{s=1}^\infty\frac{1}{e^{\beta (cs/\lambda)^2}-1} \nonumber \\
&=&\rho_0(\beta, \rho)+\rho_c(\beta) \ .
\end{eqnarray}
As in (\ref{mu-ZL}) this implies
\begin{equation}\label{mu-ZL-II}
\mu_L(\beta,\rho)=E_{1,1}-\frac{A(\beta,\rho)}{\beta L}+O(1/L^2) \ ,
\end{equation}
see (\ref{energy-ZL}), where by (\ref{dens-hierarch-II-lim}) the coefficient $A(\beta,\rho)\geq 0$
satisfies the equation
\begin{equation}\label{eq-A}
\rho = \sum_{s=1}^\infty \ \frac{1}{\beta\lambda c^2(s^2 - 1) + A} + \rho_c(\beta) \ .
\end{equation}
Hence, for $\rho>\rho_c(\beta)$ the BEC
\begin{equation*}
\rho_0(\beta, \rho)= \sum_{s=1}^\infty \ \frac{1}{\beta\lambda c^2(s^2 - 1) + A(\beta,\rho)}
\end{equation*}
is splitted between \textit{infinitely} many states in the largest
interval $L_1$, i.e. this is condensation of the \textit{type II}, \cite{VDB-Lew-Pu}.
\smallskip

\noindent \textbf{Type III BEC.} Now we show that (unusual)
\textit{spatially} fragmented \textit{type III} BEC is possible in
our hierarchical model. To split BEC between \textit{infinitely}
many states in \textit{different} intervals, let volume $\Lambda$ be
occupied by $[\ln (n +1)]$ identical (\textit{logarithmically})
large intervals:
\begin{equation}\label{L-ZL-II-inf}
L_j=\frac{\ln(\lambda L)}{\lambda} \ \ , \ \ 1\leq j \leq [\ln (n +1)]=:M_n \ \ \ {\rm{and}} \ \ \ \
L_{j > M_n}={\tilde{L}_n}:=\frac{L-L_1 M_n}{n - M_n} \ ,
\end{equation}
for $M_n < j \leq n$, of small intervals, cf. (\ref{M}). Then
(similar to (\ref{ro-M})) we get for the particle density
\begin{eqnarray}\label{ro-M-inf}
\rho_L(\beta,\mu)=\frac{1}{L}\sum_{j=1}^{M_n} \
\sum_{s=1}^\infty\frac{1}{e^{\beta(c^2s^2/L_j^2-\mu)}-1}+
\frac{n-M_n}{L}\sum_{s=1}^\infty\frac{1}{e^{\beta(c^2s^2/{\tilde{L}_n}^2-\mu)}-1} \ .
\end{eqnarray}
Since by (\ref{L-ZL-II-inf}) we have $\lim_{L \rightarrow \infty}{\tilde{L}_n}=
\lim_{L \rightarrow \infty}{(n-M_n)}/{L} = \lambda$, then the critical density (\ref{ro-crit-M=1})
rests the same. If $\rho>\rho_c(\beta)$, the equation $\rho = \rho_L(\beta,\mu)$ (\ref{ro-M-inf})
yields for asymptotics of the solution $\mu_L(\beta,\rho)$:
\begin{equation}\label{mu-ZL-M-inf}
\mu_L(\beta,\rho)=E_{j,1}-\frac{M_n}{\beta\rho_0(\beta, \rho)L}+O(1/L^2) \ , \ 1 \leq j \leq M_n \ .
\end{equation}
Then we obtain for the particle density in large intervals
\begin{eqnarray}
\nonumber
\lim_{L \rightarrow \infty}\frac{1}{L}\sum_{j=1}^{M_n} \
\sum_{s=1}^\infty\frac{1}{e^{\beta(c^2s^2/L_j^2-\mu_L(\beta,\rho))}-1}&=&
\lim_{L \rightarrow \infty}\frac{M_n}{L} \frac{1}{e^{{M_n}/{(\rho_0(\beta, \rho)L)}-O(1/L^2)}-1} \\
+ \lim_{L \rightarrow \infty}\frac{M_n}{L}
\sum_{s=2}^\infty\frac{1}{e^{\beta((\lambda c s /\ln n)^2-\mu_L(\beta,\rho))}-1}
&=& \rho_0(\beta, \rho) \ .
\label{lim-crit-inf}
\end{eqnarray}
The limit (\ref{lim-crit-inf}) and (\ref{ro-M-inf}),
(\ref{mu-ZL-M-inf}) imply that the condensate density $\rho_0(\beta,
\rho)= \rho - \rho_c(\beta)$ is splitted between  ground states of
\textit{infinitely} many \textit{logarithmic} intervals in such a
way that the condensate density in each interval is \textit{zero}.
This is a \textit{spatially} fragmented  \textit{type III} BEC,
which is different from the that corresponding to spread out over
infinite number of states discussed, e.g., in  \cite{VDB-Lew-Pu,
ZagBru}.

\smallskip
\noindent \textbf{6.3 Statistics of large Poisson intervals}

\smallskip

\noindent By virtue of \textbf{6.2} to discriminate between possible types of BEC in
the Luttinger-Sy model one has to study statistics of the size
$\left\{L_{j}^{\omega}\right\}_{j}$ of intervals induced by Poisson distributed point
impurities, see Proposition \ref{distributions}.

In fact, the first attempt to elucidated this question is already
contained in \cite{Lutt-Sy-2}. They gave some arguments in favour of
that for large finite $\Lambda$ the \textit{largest} interval
$I_{1}^{\omega}$ has a \textit{typical} length of the
\textit{logarithmic} order:
\begin{equation}\label{stat-L1}
L_{1}^{\omega} \sim \lambda^{-1}\ln(\lambda L) \ , \ L \rightarrow \infty \ .
\end{equation}
Moreover, along the same line of reasoning they conclude that all
other intervals $I_{j > 1}^{\omega}$ are typically much smaller than
$I_{1}^{\omega}$. Then neglecting the fluctuations the length of
$L_{1}^{\omega}$ they conclude that BEC has to follow  scenario we
described in \textbf{6.2} as \textit{type I} BEC, cf. (\ref{L-ZL})
and (\ref{stat-L1}).

To check these arguments and to bolster them by some rigorous reasonings we use
Proposition \ref{distributions}. First we note that the average length of the Poisson intervals is
\begin{equation}\label{exp-interv}
\mathbb{E}_{\sigma_\lambda}(L_{j_s}^{\omega})=
\lambda \int_{0}^{\infty} dL \, L \, e^{-\lambda L}= \lambda^{-1} \ \ ,
\end{equation}
i.e., the total average length of any sample of intervals
$\left\{I_{j}^{\omega}\right\}_{j=1}^{k}$ is  $k/\lambda$.
We are interested into density of the joint probability distributions generated by the events:
$\{\omega\in \Omega: L_{j_1}^{\omega}\geq L_{j_2}^{\omega}\geq \ldots \geq L_{j_k}^{\omega}\}$. They
have evidently the form
\begin{equation}\label{ordered-dens}
d{\sigma}_{\lambda, k}^{>}(L_{j_1},\ldots,L_{j_k}):=
k! \ \theta(L_{j_1}-L_{j_2})\, \theta(L_{j_2}-L_{j_3})...\theta(L_{j_{k-1}}-L_{j_k})
\ d{\sigma}_{\lambda, k}(L_{j_1},\ldots,L_{j_k}) \ .
\end{equation}
Then one gets for the joint probability density of the two \textit{largest} intervals:
\begin{equation}\label{2-largest-dens}
d{\sigma}_{\lambda, k}^{>}(L_{j_1}, L_{j_2})/dL_{j_1} dL_{j_2} = k (k-1)\, \lambda^2 \, e^{-\lambda
L_{j_1}} e^{-\lambda
L_{j_2}} \left(1-e^{-\lambda L_{j_2}}\right)^{k-2} \, \theta(L_{j_1}-L_{j_2}) \  .
\end{equation}

Now, let $A(s,t)$ be defined for non-negative integers $s$ and $t$ by
\begin{eqnarray}\label{A-funct}
A(s,t):= \int_0^\infty dx \ln^s(x)x^te^{-x} \ .
\end{eqnarray}
For $s\geq1$ and $t\geq1$ this function  verifies the following identities:
\begin{eqnarray}\label{A-relation}
A(s,t)=tA(s,t-1)+ s A(s-1,t-1)\ , \ A(1,t)=tA(1,t-1)+\Gamma(t) \ ,
\end{eqnarray}
where $\Gamma(t)$ stands for the Gamma-function. Then by
(\ref{2-largest-dens}) and (\ref{A-funct}), (\ref{A-relation})
we obtain for expectations of the two largest intervals:
\begin{align}
& \mathbb{E}_{\sigma_{\lambda, k}^{>}} (L_{j_{1}}^\omega) =
\frac{A(1, k)}{k! \ \lambda}-\frac{A(1,0)}{\lambda} =
\frac{1}{\lambda}\sum_{s=1}^k \frac{1}{s}\ ,\label{meanL1}\\
&\mathbb{E}_{\sigma_{\lambda, k}^{>}} ( L_{j_{2}}^\omega) =
\frac{A(1, k)}{k! \ \lambda}-\frac{A(1,1)}{\lambda} =
\frac{1}{\lambda}\sum_{s=2}^k \frac{1}{s}\label{meanL2} \ .
\end{align}
By virtue of (\ref{meanL1}) and (\ref{meanL2}) the mean difference
$\mathbb{E}_{\sigma_{\lambda, k}^{>}} (L_{j_{1}}^\omega- L_{j_{2}}^\omega) = 1/\lambda$ is
\textit{independent} of the number $k$ of intervals in the sample, whereas they have, for large $k$,
the \textit{logarithmic} size (cf. (\ref{stat-L1})):
\begin{equation}\label{asympt-ln-k}
\mathbb{E}_{\sigma_{\lambda, k}^{>}} (L_{j_{\, 1,2}}^\omega) =
\frac{1}{\lambda} \ln (k) + \frac{1}{\lambda} P_{1,2} + O(1/k) \ ,
\end{equation}
with respect the total average sample length $k/\lambda$, here
${P_1}= \textbf{C}:= 0,577\ldots$, is the \textit{Euler constant}, and $P_2=
\textbf{C}-1$. Using (\ref{ordered-dens}) and (\ref{A-funct}),
(\ref{A-relation}) we find that the variance of the difference
between two largest intervals in the sample is also
$k$-\textit{independent} and has the form:
\begin{equation}\label{var-diff}
Var_{\sigma_{\lambda, k}^{>}}(L_{j_{1}}^\omega- L_{j_{2}}^\omega)= \frac{1}{\lambda^2} \ .
\end{equation}
Moreover, by the joint probability distribution (\ref{2-largest-dens}) we obtain for any $\delta > 0$ that
probability
\begin{equation}\label{Pr-dist}
\mathbb{P}\{\omega : L_{j_{1}}^\omega - L_{j_{2}}^\omega > \delta \}= e^{-\lambda \delta} \
\end{equation}
of the events $A_k (\delta)= \{\omega : L_{j_{1}}^\omega - L_{j_{2}}^\omega > \delta\}$, is independent of $k$
for increasing  sequence of samples $\left\{I_{j}^{\omega}\right\}_{j=1}^{k}$, when $k \rightarrow \infty $.

By \textbf{6.2} and (\ref{asympt-ln-k}), (\ref{var-diff}) we see
that the \textit{type II} or \textit{III} BEC are impossible in the
one largest \textit{logarithmic} "box", since that total average
length of the sample is $k/\lambda$. To exclude the \textit{type
I,II,III} condensations via a \textit{space} fragmentation between,
e.g., two "boxes", we have to estimate probability of events
corresponding to the state-energy spacings between two largest
intervals. By \textbf{6.2} (see (\ref{mu-ZL-M}), (\ref{mu-ZL-II}),
(\ref{mu-ZL-M-inf}) and \cite{LPZ}) this spacing should be larger
than \textit{inverse} of the total sample length, which is
$(k/\lambda)^{-1}$. To this end it is sufficient to estimate the
probability of the event $S_{k}(a>0, \gamma > 0)$ corresponding the
spacing between ground states:
\begin{equation}\label{energy-spacing}
\mathbb{P}\{S_{k}(a,\gamma)\}:= \mathbb{P}\{\omega:
E_{s=1}(L_{j_{2}}^\omega(k))-E_{s=1}(L_{j_{1}}^\omega(k)) >
\frac{a}{k^{1-\gamma}}\} \ .
\end{equation}
Here we denote the energies in the sample $\left\{I_{j}^{\omega}\right\}_{j=1}^{k}$ by
\begin{equation}\label{energy-sample}
E_{s}(L_{j_r}^\omega(k))= \frac{c^2 s^2}{(L_{j_r}^\omega(k))^2} \ , \ r=1,...,k \ , \ s=1,2,... \ .
\end{equation}
Notice that by (\ref{Pr-dist}) we obtain that there is a kind "repulsion" between energy levels in different
intervals. Indeed,
\begin{eqnarray}
&& \mathbb{P}\{S_{k}(a,\gamma)\} \geq
\mathbb{P}\{\omega: L_{j_{1}}^\omega(k)- L_{j_{2}}^\omega(k) >
\frac{a}{2 c^2 k^{1-\gamma}}L_{j_{1}}^\omega(k)(L_{j_{2}}^\omega(k))^2\} = \nonumber \\
&&\int_{0}^\infty\int_{0}^\infty \  d{\sigma}_{\lambda, k}^{>}(x,y)
\theta (x-y -\frac{a}{2 c^2 k^{1-\gamma}} x y^2) =:p_{k}(a,\gamma) \
, \label{est-energ-space}
\end{eqnarray}
where $\lim_{k\rightarrow \infty} p_{k}(a,0<\gamma<1)=1$ by explicit
calculations in (\ref{est-energ-space}). The same argument is valid
for other than ground states as well as for intervals
$\left\{I_{j_r}^{\omega}\right\}_{r>2}^{k}$ instead of
$I_{j_2}^{\omega}$. Therefore, in this limit with the probability
$1$ the spacing is too large for fragmentation of condensate between
the largest and other intervals.

\smallskip
\noindent \textbf{6.4 The Kac-Luttinger conjecture}

\smallskip

The above arguments prove the Kac-Luttinger \textit{conjecture} in the case of the one-dimensional
random Poisson potential
of point impurities: for PBG the BEC is of \textit{type I} and it is \textit{localized} in
one "largest box".

To make this statement more precise recall that BEC exists only in the thermodynamic limit, which
we construct as \textit{increasing family} of samples of intervals $\left\{I_{j}^{\omega}\right\}_{j=1}^{k}$
induced by the point
impurities on  $\mathbb{R}$. Since these random variables are independent, we can choose increasing
sequence of independent samples with one largest interval and with the property that
\begin{equation}\label{Pr-infinity}
 \lim_{k\rightarrow \infty} \ \ \sum_{r=1}^k \mathbb{P}\{S_{k}(a,\gamma)\} = \infty \ ,
\end{equation}
see (\ref{est-energ-space}). Then by the Borel-Cantelli lemma \cite{Sh}
\begin{equation}\label{Bor-Cant}
\mathbb{P} \ \{\overline{\lim} \  S_{k}(a,\gamma)\} =1 \ ,
\end{equation}
where the event
\begin{equation*}
\overline{\lim} \  S_{k}(a,\gamma):= \bigcap_{k=1}^{\infty}\bigcup_{l=k}S_{l}(a,\gamma)
\end{equation*}
means that infinitely many events $\{S_{k}(a,\gamma)\}_{k\geq 1}$ take place. Together with \textbf{6.3}
the statement (\ref{Bor-Cant}) mean that with probability $1$ in the thermodynamic limit $\mathbb{R}$
the BEC is localized in a single "largest box", and this condensation is of the \textit{type} I.

\bigskip

\noindent \textbf{Acknowledgements}

\smallskip

\noindent V.A.Z whishes to thank Tony Dorlas (Dublin IAS), Joe Pul\'{e} (UC Dublin) and Herv\'{e} Kunz, Nicolas
Macris, Philippe Martin, Charles-Edouard Pfister (all EPF de Lausanne)
for useful discussions of different aspects of the problem considered in this paper.


\end{document}